\documentclass[10pt]{article}
\usepackage{graphicx}
\usepackage{amssymb}
\usepackage{epstopdf}
\usepackage{enumitem}
\usepackage[super,sort&compress,comma,numbers]{natbib} 
\usepackage{upgreek}
\usepackage{amsmath}
\usepackage{natmove}
\usepackage{hyperref}
\usepackage{booktabs}
\usepackage{multirow}
\hypersetup{
  colorlinks   = true,    
  urlcolor     = blue,    
  linkcolor    = blue,    
  citecolor    = red      
}
\usepackage{caption} 

\usepackage{microtype}

\title{\vspace{-2cm}\textbf{Placing Marangoni instabilities under arrest}} 
\author{M. Saad Bhamla$^\ast$$^{1}$ and Gerald G. Fuller$^{2}$\\
\normalsize{$^{1}$Department of Bioengineering},
\normalsize{$^{2}$Department of Chemical Engineering}\\
\normalsize{Stanford University, Stanford, CA 94305}\\
\normalsize{$^\ast$Corresponding author; E-mail: bhamla@stanford.edu }
}
\date{}
\begin{document}
\maketitle

\begin{figure*}[b!]
\centering
\includegraphics[width=.85\textwidth]{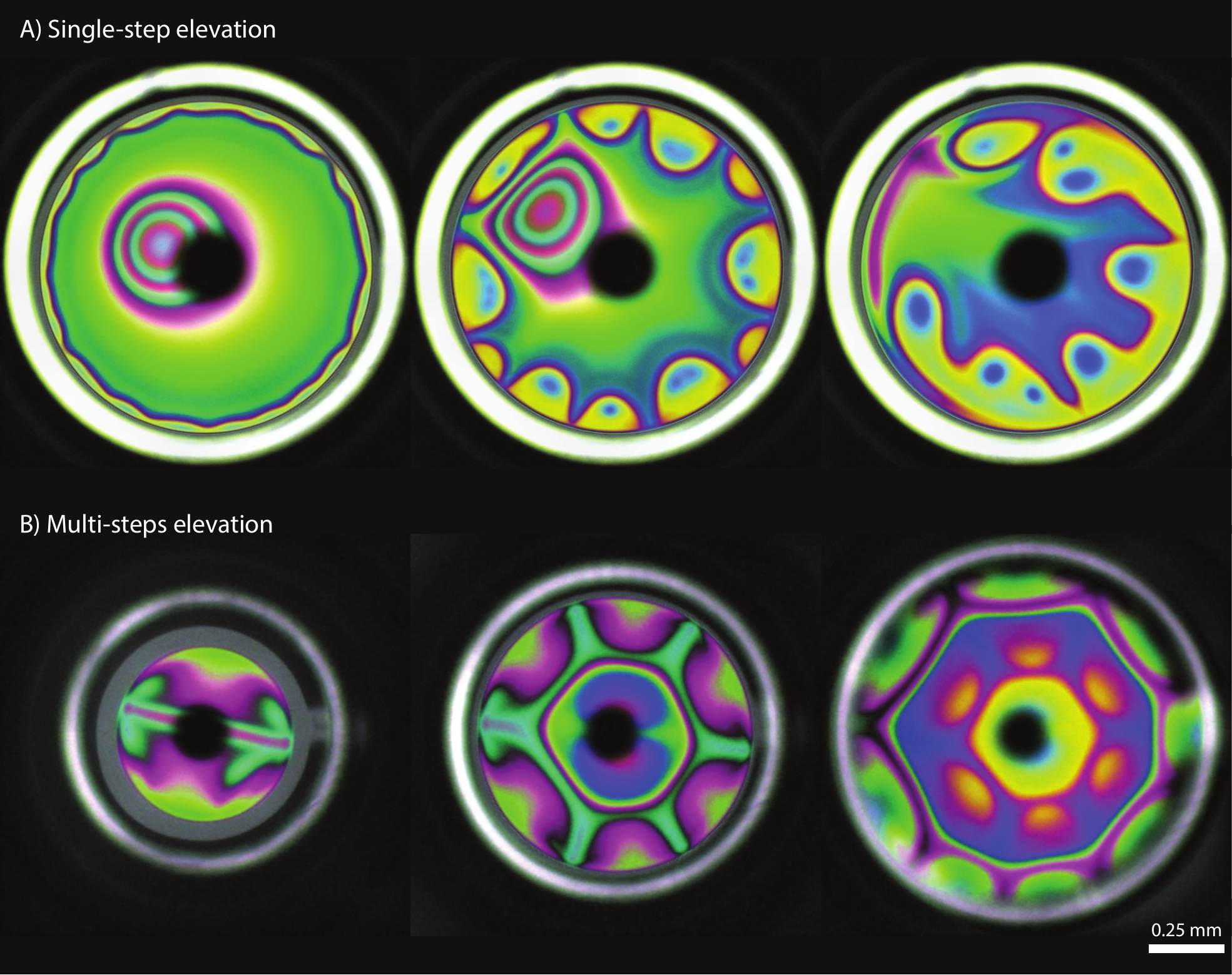}
\caption{ \label{F:Fig1}\small{Snapshots of Marangoni instabilities on a soap film elevated in A) single-step (0.9~mm) and, B) multi-steps (3 steps of 0.3~mm). In the first row, the dimple in the center is unstable and `flees' towards the periphery as plumes of surfactant rise from the periphery resulting in chaos. However, in the second row, elevation in smaller steps creates a cascade of instabilities that `arrest' each other, resulting in beautiful dynamic structures. To fully appreciate these incredible flow patterns, we urge the reader to watch the accompanying multimedia video located here http://dx.doi.org/10.1103/APS.DFD.2015.GFM.V0040. }}
\end{figure*}

Soap bubbles occupy the rare position of delighting and fascinating both young children and scientific minds alike. Sir Isaac Newton\cite{newton1979opticks}, Joseph Plateau\cite{Plateau:1873ux}, Carlo Marangoni\cite{isenberg1978science}, and Pierre-Gilles de Gennes\cite{DeGennes:2004tta}, not to mention countless others\cite{boys1958soap}, have discovered remarkable results in optics, molecular forces and fluid dynamics from investigating this seemingly simple system. We present here a compilation of curiosity-driven experiments that systematically investigate the surface flows on a rising soap bubble. From childhood experience, we are familiar with the vibrant colors and mesmerizing display of chaotic flows on the surface of a soap bubble. These flows arise due to surface tension gradients, also known as Marangoni flows or instabilities. In Figure~\ref{F:Fig1}, we show the surprising effect of layering multiple instabilities on top of each other, highlighting that unexpected new phenomena are still waiting to be discovered, even in the simple soap bubble.

\begin{figure*}[h!]
\centering
\includegraphics[width=.55\textwidth]{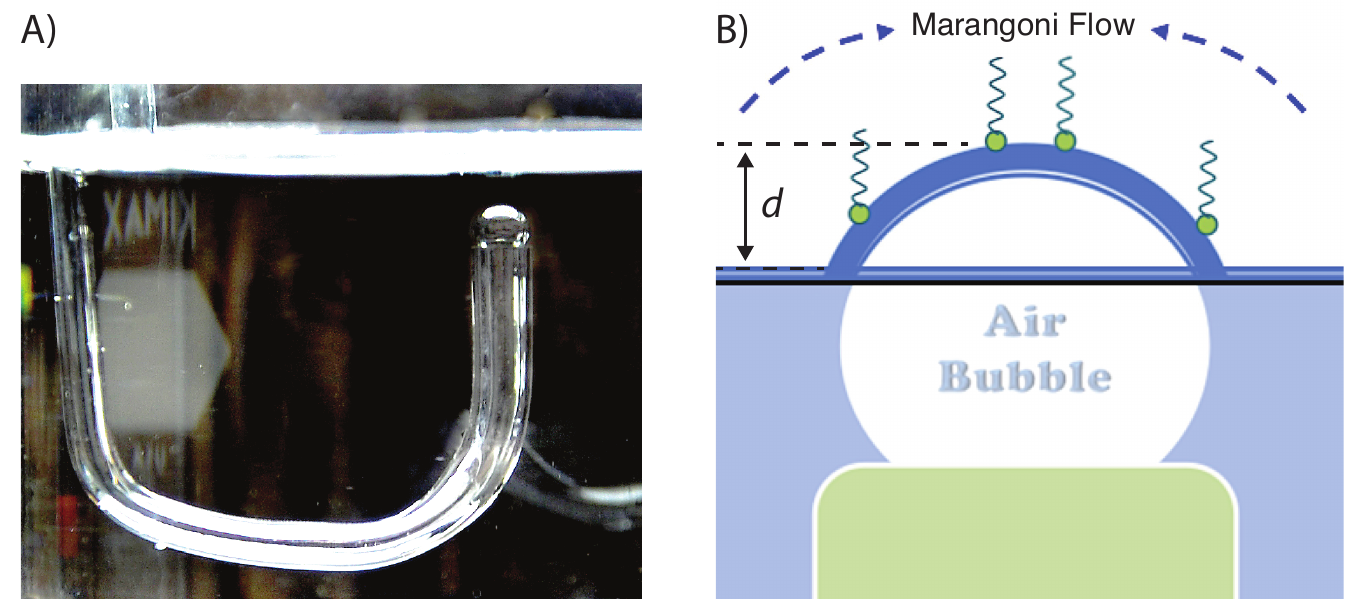}
\caption{ \label{F:Fig2} A) \small{Image of experimental setup. B) Schematic illustrating the Marangoni flow generated due to re-distribution of surfactant molecules as the bubble is elevated through the solution interface, where $d=0.9$~mm is the vertical displacement from initial starting point just beneath the air-solution interface.   }}
\end{figure*}

We conduct the experiment as follows. We place a plastic chamber containing surfactant solution (0.3~mM sodium dodecyl sulfate (SDS) in water) onto a motorized stage. In this solution, we generate an air-bubble (1~mm diameter) at the tip of a u-shaped glass capillary (Figure~\ref{F:Fig2}). We initially position the bubble just beneath the surface. Next, using the stage, we elevate the bubble in controlled, discrete steps to expose a spherical cap of the bubble at the surface. This protruding cap is essentially a draining curved film, which under white light illumination exhibits interference patterns that we record using a color CCD camera.

We find that the Marangoni instabilities can be reproduced by a broad range of soluble surfactants that have negligible surface shear viscosity. Thus, this experiment serves as a facile technique for probing the surface mobility of surfactant systems,~\cite{Bhamla:2016uy} including artificial lung surfactant therapeutics~\cite{2015SMat...11.8048H}. However, a complete theoretical analysis of this problem remains an open challenge.

\section*{Acknowledgements}
We graciously acknowledge Marco A. \`{A}lvarez-Valenzuela and John Frostad for technical support and Yevgenya Strakovsky for video editing.

\bibliographystyle{unsrt}
\bibliography{Marangoni_bhama_bibtex.bib}

\begin{thebibliography}{1}

\bibitem{newton1979opticks}
Isaac Newton.
\newblock {\em {Opticks, or, a treatise of the reflections, refractions,
  inflections {\&} colours of light}}.
\newblock Courier Corporation, 1979.

\bibitem{Plateau:1873ux}
Joseph Plateau.
\newblock {Experimental and theoretical statics of liquids subject to molecular
  forces only}.
\newblock {\em Gauthier-Villars, Paris}, 1873.

\bibitem{isenberg1978science}
Cyril Isenberg.
\newblock {\em {The science of soap films and soap bubbles}}.
\newblock Courier Corporation, 1978.

\bibitem{DeGennes:2004tta}
Pierre-Gilles De~Gennes, Fran{\c c}oise Brochard-Wyart, and David
  Qu{\'e}r{\'e}.
\newblock {\em {Capillarity and Wetting Phenomena}}.
\newblock Drops, Bubbles, Pearls, Waves. Springer Science {\&} Business Media,
  2004.

\bibitem{boys1958soap}
Charles~Vernon Boys.
\newblock {\em {Soap bubbles, their colours and the forces which mold them}},
  volume 542.
\newblock Courier Corporation, 1958.

\bibitem{Bhamla:2016uy}
M~Saad Bhamla, Chew Chai, Marco~A {\'A}lvarez-Valenzuela, Javier Tajuelo, and
  Gerald~G Fuller.
\newblock {Interfacial mechanisms for stability of surfactant-laden films}.
\newblock {\em arXiv.org}, August 2016.

\bibitem{2015SMat...11.8048H}
Eline Hermans, M~Saad Bhamla, Peter Kao, Gerald~G Fuller, and Jan Vermant.
\newblock {Lung surfactants and different contributions to thin film
  stability.}
\newblock {\em Soft Matter}, 11(41):8048--8057, October 2015.

\end{thebibliography}

\end{document}